\documentclass[a4paper,12pt]{article}
\usepackage{epsfig}
\usepackage[dvips,usenames]{color} 
\usepackage{axodraw} 
\usepackage{graphicx}

\newlength{\dinwidth} 
\newlength{\dinmargin} 
\begin{document} 
\title{$B\to PV$ Decays in the QCD Improved Factorization Approach} 
\author{Mao-Zhi Yang$^a$, Ya-Dong Yang$^b$\\ 
\small{$a$ Physics Department, Hiroshima University, Higashi-Hiroshima,  
Hiroshima 739-8526, Japan}\\ 
\small{$b$ Physics Department, Ochanomizu University, 2-1-1 Otsuka, Bunkyo-ku,  
Tokyo 112-8610, Japan}\\ 
} 
\maketitle 
 
\begin{picture}(0,0) 
       \put(335,250){HUPD-0008} 
       \put(335,270) {OCHA-PP-161} 
       \put(335,290){\bf hep-ph/0007038} 
\end{picture} 
 
\begin{abstract} 
Motivated by recent CELO  measurements and the progress of the theory of B decays,
we investigate  
$B\to P V(P=\pi,K. \,\, V= K^{*}, \rho,\omega)$  
decay modes in the framework of QCD improved factorization.  
We find that all the measured branching ratios are  
well accommodated in the reasonable parameter space and predictions  
for the other decay modes are well below the experimental upper limits.  
We also have  calculated $CP$ asymmetries in these decay modes.
 
\end{abstract} 
\newpage
 
\section{Introduction} 

It is of great interest and importance to investigate the decays of 
B mesons to charmless final states to study the weak interactions and
CP violation. In past years, we have witnessed many experimental and 
theoretical progresses in the study of B physics with the observations 
of many B charmless decay processes and improvements of the theory of
B decays.
    
Most of the theoretical studies of B decays to pseudocalar and vector
final states are based on the popular $Naive$ $Factorization$ 
approch\cite{BSW}. As it was ponited out years ago in Ref.\cite{cz},
the dominant contribution in B decays comes from the so-called
$Feynman$ $mechanism$, where the energetic quark created in the 
weak decay picks up the soft spectator softly and carries nearly all of 
the final-state meson's momentum. It is also shown that 
Pion form factor in QCD at intermediate engery scale is dominated by 
Feynman mechanism\cite{isg, rad, stef}.
From this point, we can understand why the naive factorization approch 
have worked well for B and D decays, and the many existing predictions for 
B decays based on naive factorization and spectator ansatz do have taken
in the dominant physics effects although there are shortcommings. However,
with the many new data available from CLEO and an abundance of data to 
arrive within few years from the B factories BaBar and Belle, it is demanded
highly to go beyond the naive factorization approach.

Recently, Beneke $et.\, al.,$ have formed an interesting QCD factorization formula
for B exclusive nonleptonic decays\cite{BBNS, BBNS2}. The factorization 
formula incorporates elements of the naive factorization approach(as leading 
contribution) and the hard-scattering approach(as subleading corrections), 
which allows us to calculate sysmatically radiative(subleading nonfactorizable)
 corrections  to naive factorization for B exclusive nonleptonic decays. 
An important product of the formula is that the strong final-state interaction
phases are calculable from the first principle which arise from the hard-scattering
kernel and hence process dependent. The strong phases are very important for 
studying CP violation in B decays.
Detail proofs and arguments could be found
in\cite{BBNS2}. Here we recall briefly the essence of the QCD factorization
formula as follows.    

The amplitude of $B$ decays to two light mesons, say $M_1$ and $M_2$, 
is obtained through the hadronic matrix element  
$\langle M_{1}(p_{1}) M_{2}(p_{2}){\mid}{\cal O}_i {\mid}B(p)\rangle$, 
here $M_{1}$ denotes the final meson that picks up the light spectator 
quark in the $B$ meson, and $M_{2}$ is the another meson which is composed 
of the quarks produced from the weak decay point of $b$ quark. Since the 
quark pair, forming $M_2$, is ejected from the decay point of $b$ quark 
carrying the large energy of order of $m_b$, soft gluons with the  
momentum of order of $\Lambda_{QCD}$ decouple from it at leading order 
of $\Lambda_{QCD}/m_b$ in the heavy quark limit. As a consequence any 
interaction between the quarks of $M_2$ and the quarks out of $M_2$ is 
hard at leading power in the heavy quark expansion. On the other hand, 
the light spectator quark carries the momentum of the order of $\Lambda_{QCD}$, 
and is softly transfered into $M_1$ unless it undergoes a hard interacton.  
Any soft interaction between the spectator quark and other constituents in 
$B$ and $M_1$ can be obsorbed into the transition formfactor of $B\to M_1$. 
The non-factorizable contribution to $B\to M_1 M_2$ can be calculated through 
the diagrams in Fig.1. 
 
In this paper we study $B\to PV$ decays within the framework of the QCD 
improved factorization approach \cite{BBNS,BBNS2}. 
In Sec.II We present notations and calculations. In Sec.III we compare our numerical
results with 
the experimental data presented by CLEO collabaration \cite{CLEO}. 
We find that all the measured branching ratios are  
well accommodated in the reasonable parameter space and predictions  
for the other decay modes are well below their upper limits.  
We also give our predictions of direct CP asymmetries and time integrated CP 
asymmetries in the decay modes. Large direct CP vilation asymmetries are predicted for 
the decay modes $B^0 \to \pi^0 \omega,\, K^- \rho^+ ,\, K^0 \omega$, 
$B^-\to  K^- K^{*0} ,\, K^- \rho^0 ,\, \pi^0 K^{*-},\,
  K^- \omega $. The direct CP asymmetries in the observed decay modes 
$B^- \to \pi^- \rho^0$, $\pi^- \omega$ and $\bar{B}^0 \to \pi^{\pm}\rho^{\pm}$ 
are predicted to be around few percentages.

\section{Calculations}
The effective Hamiltonian for $B$ decays is given by \cite{buras}, 
\begin{equation} 
\label{heff} 
{\cal H}_{eff} 
=\frac{G_{F}}{\sqrt{2}} \left[ V_{ub} V_{uq}^* 
\left(\sum_{i=1}^{2} 
C_{i}O_{i}^{u}+   
\sum_{i=3}^{10} 
C_{i} \, O_i + C_g O_g \right)+  
V_{cb} V_{cq}^* 
\left(\sum_{i=1}^{2} 
C_{i}O_{i}^{c}+\sum_{i=3}^{10} 
C_{i}O_{i}+C_{g}O_g \right) 
 \right], 
\end{equation} 
with the effective operators given by 
\begin{equation}\begin{array}{llllll} 
O_1^{u} & = &(\bar q_\alpha  u_\alpha)_{V-A}\cdot (\bar  
u_\beta  b_\beta)_{V-A} , 
&O_2^{u} & = & ( \bar q_\alpha  u_\beta))_{V-A}\cdot( \bar  
u_\beta  b_\alpha))_{V-A} , \\ 
O_1^{c} & = &(\bar q_\alpha  c_\alpha)_{V-A}\cdot (\bar  
c_\beta  b_\beta))_{V-A} , 
&O_2^{c} & = & ( \bar q_\alpha  c_\beta)_{V-A}\cdot (\bar  
c_\beta  b_\alpha )_{V-A}, \\ 
O_3 & = & (\bar q_\alpha  b_\alpha)_{V-A}\cdot \sum_{q'}(\bar 
 q_\beta'  q_\beta' )_{V-A},   & 
O_4 & = & (\bar q_\alpha  b_\beta))_{V-A}\cdot \sum_{q'}(\bar  
q_\beta'  q_\alpha' )_{V-A}, \\ 
O_5 & = &( \bar q_\alpha  b_\alpha)_{V-A}\cdot \sum_{q'}(\bar  
q_\beta'  q_\beta')_{V+A} ,   & 
O_6 & = & (\bar q_\alpha  b_\beta)_{V-A}\cdot \sum_{q'}(\bar  
q_\beta'  q_\alpha')_{V+A} , \\ 
O_7 & = & \frac{3}{2}(\bar q_\alpha  b_\alpha)_{V-A}\cdot  
\sum_{q'}e_{q'}(\bar q_\beta'  q_\beta' )_{V+A},   & 
O_8 & = & \frac{3}{2}(\bar q_\alpha  b_\beta)_{V-A}\cdot  
\sum_{q'}e_{q'}(\bar q_\beta'  q_\alpha')_{V+A} , \\ 
O_9 & = & \frac{3}{2}(\bar q_\alpha  b_\alpha)_{V-A}\cdot  
\sum_{q'}e_{q'}(\bar q_\beta'  q_\beta')_{V-A} ,   & 
O_{10} & = & \frac{3}{2}(\bar q_\alpha  b_\beta)_{V-A}\cdot  
\sum_{q'}e_{q'}(\bar q_\beta'  q_\alpha'\ )_{V-A},\\ 
O_g &=& (g_s/8\pi^{2}) \, m_b \, \bar{d}_{\alpha} \, \sigma^{\mu \nu} 
      \, R  \, (\lambda^A_{\alpha \beta}/2) \,b_{\beta} 
      \ G^A_{\mu \nu}~.  
\label{operators} 
\end{array} 
\end{equation} 
Here $q=d,s$  and $(q'\epsilon\{u,d,s,c,b\})$,  
$\alpha$ and $\beta$ are the $SU(3)$ color indices and  
$\lambda^A_{\alpha \beta}$, $A=1,...,8$ are the Gell-Mann matrices, 
and $G^A_{\mu \nu}$ denotes the gluonic field strength tensor. 
The Wilson coefficients evaluated at $\mu=m_b$ scale are\cite{buras} 
\begin{equation} 
\begin{array}{ll} 
        C_1= 1.082, & 
        C_2= -0.185,\\ 
        C_3=  0.014, & 
        C_4= -0.035,\\ 
        C_5=  0.009, & 
        C_6= -0.041,\\ 
        C_7= -0.002/137,& 
        C_8=  0.054/137,\\ 
        C_9= -1.292/137,& 
        C_{10}= -0.262/137,\\ 
        C_g=-0.143.& 
\end{array}\label{ci} 
\end{equation}

The non-factorizable contributions to $B\to M_1 M_2$ can be calculated through 
the diagrams in Fig.1. The results of our calculations are sumarized  compactly  
by the following equations  
 
\begin{eqnarray} 
{\cal T}_{p}=\frac{G_F}{\sqrt{2}}\sum_{p=u,c}^{} V_{pq}^{*}V_{pb} 
\biggl\{ 
    a_{1}^{p} (BM_{1},M_{2})(\bar{q}  u)_{V-A}\otimes (\bar{u}  b)_{V-A}\biggr.   
+a_{2}^{p}(BM_{1},M_{2})(\bar{u}  u)_{V-A}\otimes (\bar{q}  b)_{V-A}&  
 \nonumber\\ 
 +a_{3}^{p}(BM_{1},M_{2})(\bar{q'}  q')_{V-A}\otimes (\bar{q}  b)_{V-A}  
     + a_{4}^{p}(BM_{1},M_{2})(\bar{q}  q')_{V-A}\otimes (\bar{q'}  b)_{V-A}& 
 \nonumber\\ 
  +a_{5}^{p}(BM_{1},M_{2})(\bar{q'}  q')_{V+A}\otimes (\bar{q}  b)_{V-A}  
    +a_{6}^{p}(BM_{1},M_{2})(-2)(\bar{q}q')_{S+P}\otimes (\bar{q'}b)_{S-P}& 
\nonumber \\ 
   + a_{7}^{p}(BM_{1},M_{2}) \frac{3}{2}e_{q'} (\bar{q'}  q')_{V+A}\otimes  
             (\bar{q}  b)_{V-A}&   \\    
   + (-2)\left( a_{8}^{p}(BM_{1},M_{2}) \frac{3}{2}e_{q'}+a_{8a}(BM_{1},M_{2})\right)  
(\bar{q}q')_{S+P}\otimes (\bar{q'}b)_{S-P} 
&\nonumber \\ 
 +a_{9}^{p}(BM_{1},M_{2})\frac{3}{2}e_{q'}(\bar{q'}  q')_{V-A}\otimes  
            (\bar{q}  b)_{V-A} &\nonumber \\  
 \biggl. 
+\left( a_{10}^{p}(BM_{1},M_{2})\frac{3}{2}e_{q'}+a_{10a}^{p}(BM_{1},M_{2})\right) 
         (\bar{q}  q')_{V-A}\otimes (\bar{q'}  b)_{V-A}\biggr\},&\nonumber  
\label{tp} 
\end{eqnarray} 
where the symbol $\otimes$ denotes  
$\langle M_1 M_2|j_2\otimes j_1|B \rangle \equiv \langle M_2|j_2|0\rangle  
  \langle M_1|j_1|B\rangle $. $M_1$ represents the meson which picks up 
 the spectator quark through this paper. For $M_1$ is light $vector$ meson and  
$M_2$ is light $pseudoscalar$ meson, the effective $a_{i}^p$'s which contain  
next-to-leading order(NLO)  coefficients and ${\cal O}(\alpha_s )$ hard
scattering  
corrections are found to be 
\begin{eqnarray} 
 a_{1,2}^{c}(BV,P)&=&0, \quad a_{i}^{c}(BV,P)=a_{i}^{u}(BV,P),  
 i=3, 5, 7, 8, 9, 10, 8a, 10a. \nonumber \\ 
 a_{1}^{u}(BV,P)&=&C_{1}+\frac{C_{2}}{N} 
 +\frac{\alpha_{s}}{4\pi}\frac{C_F}{N}C_{2}F_{P},\nonumber \\ 
 a_{2}^{u}(BV,P)&=&C_{2}+\frac{C_{1}}{N} 
+\frac{\alpha_{s}}{4\pi}\frac{C_F}{N}C_{1}F_{P},\nonumber \\ 
 a_{3}^{u}(BV,P)&=&C_{3}+\frac{C_{4}}{N} 
+\frac{\alpha_{s}}{4\pi}\frac{C_F}{N}C_{4}F_{P}, \nonumber \\ 
 a_{4}^{p}(BV,P)&=&C_{4}+\frac{C_{3}}{N} 
+\frac{\alpha_{s}}{4\pi}\frac{C_F}{N} 
\left[C_{3}(F_{P}+G_{P}(s_{q})+G_{P}(s_{b}))+C_{1}G_{P}(s_{p})  
\right. \nonumber \\ 
& &\left. +(C_{4}+C_{6})\sum_{f=u}^{b}G_{P}(s_{f})+G_{V,g}\right],\nonumber\\ 
a_{5}^{u}(BV,P)&=&C_{5}+\frac{C_{6}}{N}+\frac{\alpha_{s}}{4\pi}\frac{C_F}{N} 
C_{6}(-F_{P}-12), 
\nonumber \\ 
a_{6}^{p}(BV,P)&=&C_{6}+\frac{C_{5}}{N} 
+\frac{\alpha_{s}}{4\pi}\frac{C_F}{N}  
\left[C_{1}G^{\prime}_{P}(s_{p})+C_{3}(G^{\prime}_{P}(s_{q}) 
+G^{\prime}_{P}(s_{b})) 
+(C_{4}+C_{6})\sum_{f=u}^{b}G^{\prime}_{P}(s_{f}) 
+G^{\prime}_{P,g}\right], \nonumber \\ 
a_{7}^{u}(BV,P)&=&C_{7}+\frac{C_{8}}{N}+\frac{\alpha_{s}} 
{4\pi}\frac{C_F}{N}C_{8}(-F_{P}-12), 
\nonumber \\ 
a_{8}^{p}(BV,P)&=&C_{8}+\frac{C_{7}}{N}, \nonumber \\ 
a_{8a}^{p}(BV,P)&=&\frac{\alpha_{s}}{4\pi}\frac{C_F}{N} 
\left[(C_{8}+C_{10})\sum_{f=u}^{b}\frac{3}{2}e_{f}G^{\prime}_{P}(s_{f}) 
+C_{9}\frac{3}{2}(e_{q}G^{\prime}_{P}(s_{q})+e_{b}G^{\prime}_{P}(s_{b})) 
\right], \nonumber \\ 
 a_{9}^{u}(BV,P)&=&C_{9}+\frac{C_{10}}{N}+\frac{\alpha_{s}}{4\pi} 
\frac{C_F}{N}C_{10}F_{P}, 
\nonumber \\ 
 a_{10}^{u}(BV,P)&=&C_{10}+\frac{C_{9}}{N}+\frac{\alpha_{s}}{4\pi}\frac{C_F}{N}C_{9}F_{P}, 
\nonumber \\ 
a_{10a}^{p}(BV,P)&=&\frac{\alpha_{s}}{4\pi}\frac{C_F}{N} 
\left[(C_{8}+C_{10})\frac{3}{2}\sum_{f=u}^{b}e_{f}G_{P}(s_{f}) 
+C_{9}\frac{3}{2}(e_{q}G_{P}(s_{q})+e_{b}G_{P}(s_{b})) 
\right],  
\label{aeff1} 
\end{eqnarray} 
where $q=d, s.\quad$  $q'=u, d, s$ and $f=u, d, s, c, b$.  $C_{F}=(N^2 -1)/(2N)$ and $N=3$  
is the  number of colors.  The internal quark mass in the penguin diagrams enters as  
$s_{f}=m_{f}^2/m_{b}^{2}$. $\bar{x}=1-x$ and $\bar{u}=1-u$.   
 
\begin{eqnarray}   
F_{P}&=&-12\ln\frac{\mu}{m_b } -18 + f_{P}^{I} + f_{P}^{II},  \\ 
f_{P}^{I}&=&\int_{0}^{1}dx g(x)\phi_{P}(x), {\hskip 15mm} 
g(x)= 3\frac{1-2x}{1-x}\ln x -3i\pi, \nonumber \\   
f_{P}^{II}&=&\frac{4\pi^2 }{N}\frac{f_{V}f_B }{ A_{0}^{V}(0) 
M_{B}^2 } \int_{0}^{1}dz\frac{\phi_{B}(z)}{z}  
\int_{0}^{1}dx\frac{\phi_{V}(x)}{x} 
 \int_{0}^{1}dy\frac{\phi_{P}(y)}{y}, \\ 
G_{P, g} &=& -\int_{0}^{1}dx \frac{2}{\bar{x}}\phi_{P}(x), \\ 
G_{P}(s_{q}) &=&  \frac{2}{3} - \frac{4}{3}\ln\frac{\mu}{m_b} +  
4\int_{0}^{1}dx \phi_{P}(x) \int_{0}^{1}du\quad u\bar{u}\ln 
\left[s_{q} -u\bar{u}\bar{x} -i\epsilon \right],\\ 
G^{\prime}_{P, g} &=& -\int_{0}^{1}dx \frac{3}{2}\phi^{0}_{P}(x)=-\frac{3}{2}, \\ 
G^{\prime}_{P}(s_{q}) &=&  \frac{1}{3} - \ln\frac{\mu}{m_b} +  
3\int_{0}^{1}dx \phi^{0}_{P}(x) \int_{0}^{1}du\quad u\bar{u}\ln 
\left[s_{q} -u\bar{u}\bar{x} -i\epsilon \right], 
\label{func1} 
\end{eqnarray}

For $M_1$ is $pseudoscalar$ and $M_2$ is $vector$, the co-efficents are 
\begin{eqnarray} 
 a_{1,2}^{c}(BP,V)&=&0, \quad a_{i}^{c}(BP,V)=a_{i}^{u}(BP,V),  
i=3, 5, 7, 8, 9, 10, 8a, 10a. \nonumber \\ 
 a_{1}^{u}(BP,V)&=&C_{1}+\frac{C_{2}}{N} 
+\frac{\alpha_{s}}{4\pi}\frac{C_F}{N}C_{2}F_{V},\nonumber \\ 
 a_{2}^{u}(BP,V)&=&C_{2}+\frac{C_{1}}{N} 
+\frac{\alpha_{s}}{4\pi}\frac{C_F}{N}C_{1}F_{V},\nonumber \\ 
 a_{3}^{u}(BP,V)&=&C_{3}+\frac{C_{4}}{N} 
+\frac{\alpha_{s}}{4\pi}\frac{C_F}{N}C_{4}F_{V}, \nonumber \\ 
 a_{4}^{p}(BP,V)&=&C_{4}+\frac{C_{3}}{N} 
+\frac{\alpha_{s}}{4\pi}\frac{C_F}{N} 
\left[C_{3}(F_{V}+G_{V}(s_{q})+G_{V}(s_{b}))+C_{1}G_{V}(s_{p})  
\right. \nonumber \\ 
& &\left. +(C_{4}+C_{6})\sum_{f=u}^{b}G_{V}(s_{f})+G_{V,g}\right],\nonumber\\ 
a_{5}^{u}(BP,V)&=&C_{5}+\frac{C_{6}}{N} 
+\frac{\alpha_{s}}{4\pi}\frac{C_F}{N}C_{6}(-F_{V}-12), 
\nonumber \\ 
a_{6}^{p}(BP,V)&=&C_{6}+\frac{C_{5}}{N}, \nonumber \\ 
a_{7}^{u}(BP,V)&=&C_{7}+\frac{C_{8}}{N} 
+\frac{\alpha_{s}}{4\pi}\frac{C_F}{N}C_{8}(-F_{V}-12), 
\nonumber \\ 
a_{8}^{p}(BP,V)&=&C_{8}+\frac{C_{7}}{N}, \nonumber \\ 
a_{9}^{u}(BP,V)&=&C_{9}+\frac{C_{10}}{N} 
+\frac{\alpha_{s}}{4\pi}\frac{C_F}{N}C_{10}F_{V}, 
\nonumber \\ 
 a_{10}^{u}(BP,V)&=&C_{10}+\frac{C_{9}}{N} 
+\frac{\alpha_{s}}{4\pi}\frac{C_F}{N}C_{9}F_{V}, 
\nonumber \\ 
a_{10a}^{p}(BP,V)&=&\frac{\alpha_{s}}{4\pi}\frac{C_F}{N} 
\left[(C_{8}+C_{10})\frac{3}{2}\sum_{f=u}^{b}e_{f}G_{V}(s_{f}) 
+C_{9}\frac{3}{2}(e_{q}G_{V}(s_{q})+e_{b}G_{V}(s_{b})) 
\right],  
\label{aeff2} 
\end{eqnarray} 
where  
\begin{eqnarray}   
F_{V}&=&-12\ln\frac{\mu}{m_b } -18 + f_{V}^{I} + f_{V}^{II},  \\ 
f_{V}^{I}&=&\int_{0}^{1}dx g(x)\phi_{V}(x), {\hskip 15mm} 
g(x)= 3\frac{1-2x}{1-x}\ln x -3i\pi, \nonumber \\   
f_{V}^{II}&=&\frac{4\pi^2 }{N}\frac{f_{P}f_B }{ f_{+}^{B\to P}(0) 
M_{B}^2 } \int_{0}^{1}dz\frac{\phi_{B}(z)}{z}  
\int_{0}^{1}dx\frac{\phi_{P}(x)}{x} 
 \int_{0}^{1}dy\frac{\phi_{V}(y)}{y}, \\ 
G_{V, g} &=& -\int_{0}^{1}dx \frac{2}{\bar{x}}\phi_{V}(x), \\ 
G_{V}(s_{q}) &=&  \frac{2}{3} - \frac{4}{3}\ln\frac{\mu}{m_b} +  
4\int_{0}^{1}dx \phi_{V}(x) \int_{0}^{1}du\quad u\bar{u}\ln 
\left[s_{q} -u\bar{u}\bar{x} -i\epsilon \right]. 
\label{func2} 
\end{eqnarray} 
Where $\phi_{P}(x)$ and $\phi^{0}_{P}(x)$ are the pseudoscalar meson's  twist-2 
and twist-3 distribution amplitudes (DA) respectively. $\phi_{V}(x)=\phi_{V,\|}$
is the leading twist DA for the longitudinally polarized vector meson states. We 
have used the fact that light vector meson is mainly longitudinally polarized in 
B decays and the tranversly polarized state is suppressed by the factor 
$M_{V}/M_{B}$. Further more in $B\to PV$ decays the vector meson should be 
completely polarized in longitudinal direction due to the angular momentum 
conservation requirement. In the derivation of the effective coefficients
$a_i$'s we have used NDR scheme and assumption of 
asymtotic DAs. The infared divergences in $Fig.1.a-d$ are cancelled in their sum.

With the effective coefficients in Eqs.(\ref{aeff1}) and (\ref{aeff2}), 
we can write down the decay amplitudes of the decay modes (we only list 
four decay modes which have been detected in experiment here. The other 
decay modes are given in appendix A) 
\begin{eqnarray} 
{\cal M}( \bar{B}^0 \to\pi^+ \rho^- )=\frac{G_{F}}{\sqrt{2}} 
 f_{\rho}M_{B}^2 F_{0}^{B\to\pi}\lambda V_{cb}  
  \biggl\{ 
            R_{u} e^{-i\gamma}  
      \left[ a_{1}(\bar{B}^0 \pi^{+},\rho^-  ) 
                     +a_{4}^{u}(\bar{B}^0 \pi^{+},\rho^-  ) 
  \right. \biggr. &\nonumber \\ 
   \biggl. \left.   +a_{10}(\bar{B}^0 \pi^{+},\rho^-  ) 
                       +a_{10a}(\bar{B}^0 \pi^{+},\rho^-  ) 
            \right] 
     -\left[ a_{4}^{c}(\bar{B}^0 \pi^{+},\rho^-  )+a_{10}(\bar{B}^0 \pi^{+},\rho^-  ) 
                +a_{10a}(\bar{B}^0 \pi^{+},\rho^-  ) 
      \right]  
    \biggr\};&  
\end{eqnarray} 
 
\begin{eqnarray} 
{\cal M}(\bar{B}^0 \to\pi^- \rho^+ )=\frac{G_{F}}{\sqrt{2}} 
 f_{\pi}M_{B}^2 A_{0}^{B\to \rho} \lambda V_{cb}  
\biggl\{ 
      R_{u} e^{-i\gamma}  
         \left[  a_{1}(\bar{B}^0 \rho^{+},\pi^- )  
     +a_{4}^{u}(\bar{B}^0 \rho^{+},\pi^-  ) 
   \right. \biggr. &\nonumber \\ 
   \left.  +\left( a_{6}^{u}(\bar{B}^0 \rho^{+},\pi^- )+a_{8}(\bar{B}^0 \rho^{+},\pi^- ) 
        +a_{8a}(\bar{B}^0 \rho^{+},\pi^- ) 
           \right) R_{\pi^- } 
+a_{10}(\bar{B}^0 \rho^{+},\pi^-  )+a_{10a}(\bar{B}^0 \rho^{+},\pi^- ) 
\right] 
&\nonumber \\ 
  -\left[ a_{4}^{c}(\bar{B}^0 \rho^{+},\pi^-  ) 
        +\left( a_{6}^{c}(\bar{B}^0 \rho^{+},\pi^-  )+a_{8}(\bar{B}^0 \rho^{+},\pi^-  ) 
            +a_{8a}(\bar{B}^0 \rho^{+},\pi^- )  
         \right) R_{\pi^- } 
    \right. &\nonumber \\  
\biggl.       
     \left. +a_{10}(\bar{B}^0 \rho^{+},\pi^-  )+a_{10a}(\bar{B}^0 \rho^{+},\pi^-  ) 
      \right] 
\biggr\};&  
\end{eqnarray} 
 
\begin{eqnarray} 
{\cal M}(B^- \to\pi^- \rho^0 )=\frac{G_{F}}{2} f_{\rho}M_{B}^2  
F_{0}^{B\to\pi}\lambda V_{cb}  
\biggl\{ 
   R_{u} e^{-i\gamma}  
\left[ 
a_{2}(B^{-}\pi,\rho^{0})-a_{4}^{u}(B^{-}\pi,\rho^{0})+\frac{3}{2} 
 \left( a_{7}(B^{-}\pi,\rho^{0}) 
 \right. \right. \biggr. &\nonumber \\ 
\left. \left. +a_{9}(B^{-}\pi,\rho^{0})\right)  
+\frac{1}{2} a_{10}(B^{-}\pi,\rho^{0})-a_{10a} \right] 
&\nonumber \\ 
\biggl. -\left[ -a_{4}^{c}(B^{-}\pi,\rho^{0})+\frac{3}{2}  
     \left( a_{7}(B^{-}\pi,\rho^{0}) 
           +a_{9}(B^{-}\pi,\rho^{0})\right) 
                +\frac{1}{2} a_{10}(B^{-}\pi,\rho^{0})-a_{10a} \right] 
                                \biggr\}  & \nonumber \\ 
+ \frac{G_{F}}{2} f_{\pi}M_{B}^2 A_{0}^{B\to \rho} \lambda V_{cb} 
\biggl\{ 
        R_{u} e^{-i\gamma} 
       \left[ a_{1}(B^{-}\rho^{0},\pi^{-})+a_{4}^{u}(B^{-}\rho^{0},\pi^{-})  
                  \right.\biggr.  &\nonumber \\ 
         \left. +\left( a_{6}^{u}(B^{-}\rho^{0},\pi^{-})+a_{8}(B^{-}\rho^{0},\pi^{-}) 
                    +a_{8a}(B^{-}\rho^{0},\pi^{-})\right) R_{\pi^- } 
                  +a_{10}(B^{-}\rho^{0},\pi^{-})+a_{10a} 
         \right] &\nonumber \\ 
  -\left[ a_{4}^{c}(B^{-}\rho^{0},\pi^{-})+ 
       \left( a_{6}^{c}(B^{-}\rho^{0},\pi^{-}) 
             +a_{8}(B^{-}\rho^{0},\pi^{-})+a_{8a}(B^{-}\rho^{0},\pi^{-})\right) R_{\pi^- } 
   \right. &\nonumber \\ 
   \biggl.\left. +a_{10}(B^{-}\rho^{0},\pi^{-})+a_{10a}(B^{-}\rho^{0},\pi^{-}) 
          \right]      
    \biggr\};& 
\end{eqnarray} 
\begin{eqnarray} 
{\cal M}(B^- \to\pi^- \omega)=\frac{G_{F}}{2} f_{\pi}M_{B}^2 A_{0}^{B\to\omega }  
\lambda V_{cb}  
\biggl\{ 
       R_{u} e^{-i\gamma}  
         \left[ a_{1}(B^- \omega,\pi^- ) 
       +a_{4}^{u}(B^- \omega,\pi^- )+\left( a_{6}^{u}(B^- \omega,\pi^- )  
       \right.\right. 
 &\nonumber \\      
      \left. \left. 
         +a_{8}(B^- \omega,\pi^- )+a_{8a}(B^- \omega,\pi^- )  
      \right) R_{\pi^- }  
          +a_{10}(B^- \omega,\pi^- )+a_{10a}(B^- \omega,\pi^- ) 
      \right] 
&\nonumber \\ 
    -\left[ a_{4}^{c}(B^- \omega,\pi^- )+\left( a_{6}^{c}(B^- \omega,\pi^- ) 
                                +a_{8}(B^- \omega,\pi^- ) 
              +a_{8a}(B^- \omega,\pi^- ) \right) R_{\pi^- } 
     \right.&\nonumber \\ 
\biggl. 
     \left.  +a_{10}(B^- \omega,\pi^- )+a_{10a}(B^- \omega,\pi^- ) 
      \right] 
\biggr\}&\nonumber \\ 
+\frac{G_{F}}{2} f_{\omega}  M_{B}^2 F_{0}^{B\to\pi}\lambda V_{cb}  
\biggl\{ 
             R_{u} e^{-i\gamma}  
        \left[ a_{2}(B^- \pi^{-}, \omega) 
                   +a_{4}^{u}(B^- \pi^{-}, \omega)  
            +2\left( a_{3}(B^- \pi^{-}, \omega)+a_{5}(B^- \pi^{-}, \omega) \right) 
        \right. 
\biggr.&\nonumber \\ 
  \left. 
   +\frac{1}{2} 
   \left( a_{7}(B^- \pi^{-}, \omega) 
        +a_{9}(B^- \pi^{-}, \omega) 
        -a_{10}(B^- \pi^{-}, \omega)+2 a_{10a}(B^- \pi^{-}, \omega) 
   \right) 
 \right] &\nonumber \\ 
-\left[  
      a_{4}^{c}(B^- \pi^{-}, \omega) 
     +2\left( a_{3}(B^- \pi^{-}, \omega)+a_{5}(B^- \pi^{-}, \omega) \right) 
+\left( a_{7}(B^- \pi^{-}, \omega) 
+a_{9}(B^- \pi^{-}, \omega) 
\right. \right. &\nonumber \\ 
\biggl. \left. \left. 
-a_{10}(B^- \pi^{-}, \omega)+2 a_{10a}(B^- \pi^{-}, \omega)\right) /2 
\right] 
\biggr\};& 
\end{eqnarray} 
 
Where $R_b =\frac{1-\lambda^{2}/2}{\lambda}{\mid} \frac{V_{ub}}{V_{cb}}{\mid}$ and 
$R_{b}^{\prime} =\frac{\lambda}{1-\lambda^{2}/2} {\mid}\frac{V_{ub}}{V_{cb}}{\mid}$.   
$V_{cb}, V_{ud}$ and $V_{us}$ are chosen to be real and $\gamma$ is the phase of  
$V^{*}_{ub}$. $\lambda=|V_{us}|=0.2196$. 
$R_{\pi^{\mp}}=-2M^{2}_{\pi^{\mp}}/(m_{b}(m_u +m_d ))$.

\section{Numerical calculations and discussions of results} 
 
In the numerical calculations we use \cite{pdg} 
$$\tau(B^+)=1.65\times 10^{-12}s, ~~~\tau(B^0)=1.56\times 10^{-12}s,$$ 
$$ M_B = 5.2792 { GeV},~~~~ m_b =4.8GeV, ~~~~m_c =1.4 GeV,$$ 
$$f_B = 0.180GeV, ~~f_\pi = 0.133{ GeV}, ~~f_K=0.158 GeV,$$ 
$$f_{K^*}=0.214GeV, ~~f_{\rho}=0.21GeV, ~~f_{\omega}=0.195GeV.$$
For the chiral enhancement factors for the pseudoscalar mesons,  
we take 
$$
R_{\pi^{\pm}}=R_{K^{\pm,0}}=-1.2,
$$
which are consistent with the values used in \cite{BBNS,chengy,hsw}. We
should take care for $R_{\pi^0}$. As pointed out in Ref.\cite{BBNS2}, 
 $R_{\pi^0}$ for $\pi^0$ should be $-2M_{\pi}^{2}/(m_{b}(m_u +m_d ))$
and equal to $R_{\pi^{\pm}}$
due to inclusion isospin breaking effects correctly.
 
For the form factors, we take the results of light-cone 
sum rule \cite{PBVB, ball} 
 
 $$F^{B\to \pi}(0)=0.3,~~F^{B\to K}(0)=1.13F^{B\to \pi}(0),$$ 
 $$A_0^{B\to\rho}=0.372, ~~A_0^{B\to K^*}=0.470,$$ 
and assume $A_0^{B\to \omega}(0)=1.2*A_0^{B\to\rho}(0)$ since we find larger 
 $A_0^{B\to \omega}(0)$ is preferred by experimental data.

We take the leading-twist DA $\phi(x)$ and the twist-3 DA $\phi^0(x)$ of 
light pseudoscalar and vector mesons as the asymptotic form \cite{wave} 
\begin{equation} 
\phi_{P,V}(x) =  6 x (1-x), \quad \phi^{0}_{P}(x) =1. 
\label{phi} 
\end{equation} 
For the $B$ meson, the wave function is chosen as \cite{KLS,LUY}, 
\begin{equation} 
\phi_B(x) = N_{B} x^2(1-x)^2 \mathrm{exp} \left 
 [ -\frac{M_B^2\ x^2}{2 \omega_{B}^2} \right], 
\label{bwav} 
\end{equation} 
with  $\omega_{B}=0.4$ GeV, and $N_{B}$ is the  
normalization constant to make $\int_{0}^{1} dx \phi_{B}(x) =1$. 
$\phi_{B}(x)$  is strongly peaked around $x=0.1$, which is  
consistent with the observation of Heavy Quark Effective Theory that the  
wave function should be peaked around $\Lambda_{QCD}/M_{B}$. 
 
We have used the unitarity of the CKM matrix  
$V_{uq}^* V_{ub} + V_{cq}^* V_{cb} + V_{tq}^* V_{tb}=0$  to decompose the amplitudes  
into terms containing $V_{uq}^* V_{ub}$ and  $V_{cq}^* V_{cb}$, and  
\begin{equation}\begin{array}{ll} 
|V_{us}|=\lambda=0.2196,& |V_{ub}/V_{cb}|=0.085\pm 0.02, \\ 
|V_{cb}|=0.0395\pm 0.0017,& |V_{ud}|=1-\lambda^{2}/2. 
\end{array} 
\end{equation} 
We leave the CKM angle $\gamma$ as a free parameter. 
 
The branching ratios of two body $B$ decays is given by 
\begin{equation} 
Br(B\to M_1 M_2)=\frac{\tau_B}{16\pi M_B }|{\cal M}(B\to M_1 M_2)|^2 . 
\end{equation} 
For the case that the final state $f$ is non-$CP$-eigenstate,  
the direct CP asymmetry parameter is defined as 
\begin{equation} 
A^{dir}_{CP}=\frac{|{\cal M}(B^+\to f)|^2-|{\cal M}(B^-\to \bar{f})|^2} 
                      {|{\cal M}(B^+\to f)|^2+|{\cal M}(B^-\to \bar{f})|^2}, 
\label{dircp1} 
\end{equation} 
and 
\begin{equation} 
A^{dir}_{CP}=\frac{|{\cal M}(B^0\to f)|^2-|{\cal M}(\bar{B}^0\to \bar{f})|^2} 
                      {|{\cal M}(B^0\to f)|^2+|{\cal M}(\bar{B}^0\to \bar{f})|^2}. 
\label{dircp2} 
\end{equation} 
For the neutral B decaying into CP eigenstate $f$, i.e., $f=\bar{f}$, 
the effects of $B^0-\bar{B^0}$ mixing should be taken 
into account. Thus the CP asymmetry is time dependent, 
which is given by\cite{Grcp} 
\begin{equation} 
A_{CP}(t)=A^{dir}_{CP}cos(\Delta mt)- 
             \frac{2Im(\xi)}{1+|\xi|^2} sin(\Delta mt), 
\label{mixcp1} 
\end{equation} 
and the time-integrated $CP$ asymmetry is obtained through 
\begin{equation} 
A_{cp}=\frac{1-|\xi|^2-2Im\xi(\Delta m/\Gamma)} 
           {(1+|\xi|^2)[1+(\Delta m/\Gamma)^2]}, 
\label{mixcp1} 
\end{equation} 
where $\Delta m$ is the mass difference of the two mass eigenstates of neutral B 
mesons, and $A^{dir}_{CP}$ is the direct CP asymmetry defined in  
eq.(\ref{dircp2}). The parameter $\xi$ is given by 
\begin{equation} 
\xi=\frac{V_{tb}^*V_{td}\langle f|H_{eff}|\bar{B^0}\rangle} 
                  {V_{tb}V_{td}^*\langle f|H_{eff}|B^0\rangle}. 
\end{equation} 
 
The numerical results of the branching ratios $B\to PV$ are shown in 
Fig.2 as the function of CKM angle $\gamma$. We can see from Fig.2.1, 
2.2, and 2.3 that for the three detected channels the predicted branching 
ratios agree  well with the CLEO experiment data \cite{CLEO}. Our predictions for
other decay modes  are well below their $90\% C.L$ upper limits.  

There are several works available with deatil analyses of the CLEO new data 
of the decays of B to charmless PV states\cite{chengy,hsw,gro}. 
Cheng and Yang have renewed their
predictons with the ``generalized factorization" framework\cite{chengy}.
It is worth to note that the shortcommings in the ``generalized factorization" 
are resolved in the framework employed in this paper. Nonfactorizable effects
are calculated in a rigorous way here instead of being parameterized by 
effective color number. Since the hard scattering kernals are convoluted
with the  light cone DAs of the mesons, gluon virtuality $k^2 ={\bar x}m_b^2$ 
in the penguin diagram Fig.1.e has well defined meaning and leaves no 
ambiguity as to the value of $k^2$, which has usually been treated as a free
phenomenological parameter in the estmations of the strong phase generated 
though the BSS mechanism\cite{BSS}. So that CP asymmetries are predicted 
soundly in this paper. We present the numerical result of the branching 
ratios of $B\to PV$ decays in Table. \ref{tab1} with the relevant strong phases
shown explicitly. It shows that the strong phases are generally mode dependent.

\begin{table}[htbp]
\caption{ Strong phases in the branching ratios (in units of $10^{-6}$) for 
the charmless decays modes studied by CLEO. $\gamma=Arg V_{ub}^*$.
}
\begin{center}
	\begin{tabular}{cc}
		\hline  
            \hline    
$B( B^{-}\to \pi^- \rho^0 )=6.65|0.11 e^{-i86.5^{\circ}}+ e^{-i\gamma}|^2$ & 
$B( \bar{B}^0  \to \pi^+ \rho^- )=19.79|0.11 e^{i9.02^{\circ}}+ e^{-i\gamma}|^2  $\\
\hline
$B( \bar{B}^0  \to \pi^- \rho^+ )=13.43|0.03 e^{i172^{\circ}}+ e^{-i\gamma}|^2$ &
$B( B^- \to \pi^- \omega)=10.59|0.065 e^{i26.01^{\circ}} + e^{-i\gamma}|^2 $\\
\hline
$B( \bar{B}^0  \to \pi^0 \rho^0 )=0.11|0.21 e^{2.90^{\circ}}+ e^{-i\gamma}|^2  $&
$B( B^- \to \pi^0 \rho^- )=10.81|0.176 e^{i7.20^{\circ}} + e^{-i\gamma}|^2 $\\
\hline
$B( \bar{B}^0  \to \pi^0 \omega)=1.49\times 10^{-3}|1.64 e^{i148^{\circ}} + 
e^{-i\gamma}|^2  $&
 $B( B^- \to K^- \rho^0 )=0.55|0.24 e^{-i162^{\circ}}+ e^{-i\gamma}|^2  $\\
\hline
$B(  B^- \to \pi^- \bar{K}^{*0})=0.0012|56.4e^{-i15.7^{\circ}}+ e^{-i\gamma}|^2 $& 
$ B( B^- \to K^- K^{*0})=0.030|2.86e^{i164^{\circ}} + e^{-i\gamma}|^2 $  \\
\hline
$B(  B^- \to \pi^0 K^{*-})=0.59|2.80e^{-i169^{\circ}}+ e^{-i\gamma}|^2  $& 
 $B(  B^- \to K^- \omega )=0.80|0.48e^{-i9.23^{\circ}}+ e^{-i\gamma}|^2 $\\
\hline
$B( \bar{B}^0  \to \bar{K^0} \omega)=0.72|0.81e^{-i11.8^{\circ}}+ e^{-i\gamma}|^2  $&
$B( \bar{B}^0  \to K^- \rho^{+})=0.96|0.63e^{-i7.20^{\circ}}+ e^{-i\gamma}|^2 $\\
\hline
$B( \bar{B}^0  \to \pi^0 \bar{K}^{*0})=.004|12.89e^{i67.61^{\circ}}+ e^{-i\gamma}|^2  $& \\
\hline
\hline
	\end{tabular}
\end{center}
\label{tab1}
\end{table}

Hou, Smith and W\"urthwein have performed a model dependent fit using
the recent CLEO data and found $\gamma=114^{+25}_{-21}$ degree. Using 
SU(3) flavor symmetry, Gronau and Rosner have analysed the decays of B 
to charmless PV final states extensivly and found several processes are 
consistent with $\cos\gamma< 0$. In this paper we find  $\cos\gamma< 0$ 
is favored by the $B^- \to \pi^- \rho^0$ and $\bar{B}^0 \to \pi^- \rho^+ +
\pi^+ \rho^- $ if their experimental center values are taken seriously. 
To meet its center value with $\cos\gamma< 0$, $B^- \to \pi^- \omega$ 
would indicate larger form factor i.e. $A_{0}^{B\to\omega}(0)> A_{0}^{B\to\rho}(0)$.
In our numerical calculation, we have taken $A_{0}^{B\to\omega}(0)=0.446$ which 
is still consistent with the LCSR results $0.372\pm 0.074$\cite{PBVB}.
It is also interesting to note that $\bar{B}^0 \to \pi^+ \rho^- $ is suppressed by
$\cos\gamma< 0$ while  $\bar{B}^0 \to \pi^- \rho^+ $ is enchanced. The defference
between $Br(\bar{B}^0 \to \pi^+ \rho^- )$   and $Br(\bar{B}^0 \to \pi^- \rho^+ )$
is much more sensitive to $\gamma$ than their sum.

For comparison with the results in the literature, we table our predictions
made for  $\gamma=100^{\circ}$ in Table.\ref{tab2}. We find that most of 
our predictions agree
 with  Ref.\cite{chengy}. For $Br(\bar{B}^{0}\to \pi^0 \bar{K}^{*0})$, our 
prediction is much smaller than the prediction of Ref.\cite{chengy} which exceeds the
upper limit slightly.    

\begin{table}[htbp]
\caption{ Branching ratios (in units of $10^{-6}$) for the charmless decays
 modes studied by CLEO. 
Experimential results are taken from \cite{CLEO}. Our results are made  for
$\gamma=100^{\circ}$. Cheng and Yang's preferred predictions\cite{chengy} 
(the case $N_{c}^{eff}(LL)=2$ and $N_{c}^{eff}(LR)=6$) are taken for comparison. 
The form factors used in Ref.\cite{chengy} are very simliar to ours.
}
\begin{center}
	\begin{tabular}{ccccc}
		\hline  
            \hline
Decay modes   & Our results & Ref\cite{chengy} & CLEO  $B_{fit}$ \cite{CLEO}  & 
$B$ or $90\% \, B$ UL\cite{CLEO}  \\
\hline
$ B^{-}\to \pi^- \rho^0 $  & 8.96  & 13.0 & $10.4^{+3.3}_{-3.4}\pm 2.1$  
& $10.4^{+3.3}_{-3.4}\pm 2.1$       \\
\hline
$\bar{B}^0  \to \pi^+ \rho^- $&18.6 &18.2 &Sum up&  Sum up\\
$\bar{B}^0  \to \pi^- \rho^+ $&13.5 &14.2 & $27.6^{+8.4}_{-7.4}\pm 4.2$ 
& $27.6^{+8.4}_{-7.4}\pm 4.2$   \\
\hline
$B^- \to \pi^- \omega $&9.82 &10.7& $11.3^{+3.3}_{-2.9}\pm 1.4$ 
& $11.3^{+3.3}_{-2.9}\pm 1.4$ \\
\hline
$\bar{B}^0  \to \pi^0 \rho^0 $&0.11 &0.75 & $1.6^{+2.0}_{-1.4}\pm 0.8$ &$<$5.5 \\
\hline
$B^- \to \pi^0 \rho^- $&10.0 &13.1 & &$<$43 \\
\hline
$\bar{B}^0  \to \pi^0 \omega $&0.004 & 0.02&$0.8^{+1.9+1.0}_{-0.8-0.8}$ &$<$5.5\\
\hline
 $B^- \to K^- \rho^0 $&0.71 &1.10 & $8.4^{+4.0}_{-3.4}\pm 1.8$ &$<$17 \\
\hline
$ B^- \to \pi^- \bar{K}^{*0}    $& 3.85 & 3.64 & $7.6^{+3.5}_{-3.0}\pm 1.6$ &$<$16 \\
\hline
$ B^- \to K^- K^{*0}$ & 0.26&0.39 & $0.0^{+1.3+0.6}_{-0.0-0.0}$ &$<$5.3\\
\hline
$ B^- \to \pi^0 K^{*-} $& 6.42& 4.34 & $7.1^{+11.4}_{-7.1}\pm 1.0$ &$<$31\\
\hline
 $ B^- \to K^- \omega $&0.97 & 2.24& $3.2^{+2.4}_{-1.9}\pm 0.8$ &$<$7.9\\
\hline
$\bar{B}^0  \to \bar{K^0} \omega $&0.40&1.89 & $10.0^{+5.4}_{-4.2}\pm1.4$ &$<$21 \\
\hline
$\bar{B}^0  \to K^- \rho^{+}$&1.29 & 3.49 & $16.0^{+7.6}_{-6.4}\pm 2.8$&$<$32 \\
\hline
$\bar{B}^0  \to \pi^0 \bar{K}^{*0} $&0.58 & 3.92& $0.0^{+1.3+0.5}_{-0.0-0.6}$&$<$3.6 \\
\hline
\hline
	\end{tabular}
\end{center}
\label{tab2}
\end{table}

The direct and time-integrated $CP$ asymmetries are shown in Fig3. and  
Fig.4 respectively. 
For $\gamma$ around $100^{\circ}$, the direct CP asymmetries in the decay
modes $B^- \to  K^- K^{*0}$, $K^-\rho^0$,
$\pi^0 K^{*-}$, $K^{-}\omega$ and $\bar{B}^{0}\to K^- \rho^+ $, 
$\bar{K}^0 \omega$ 
are as large as $(\pm)10\% \sim (\pm)20\%$. Unfortunatly the decay rates of 
these decay modes are quite small. From Fig.4, we can see that the time 
integrated CP asymmetries in $\bar{B}^0 \to \rho^+ \pi^- $, $\rho^0 \pi^0$ , 
$\omega\pi^0 $
are about $\pm 30\%\sim \pm 60\%$ for $\gamma$ around $100^{\circ}$.

\section{Summary} 
 
In this paper we have calculated the branching ratios and CP asymmetries of the 
charmless decays $B\to PV(P=(\pi, K), V=(\rho, \omega,K^* ))$ in the QCD improved 
factorization approach which have been formed recently by Beneke $et.\, al.$
\cite{BBNS,BBNS2}.

We have used LCSR form factors $F^{B\to\pi,K}(0)$ and $A_{0}^{\rho,K^* }(0)$ as
inputs. The results of $Br(B^- \to\pi^- \rho^0)$ and  
$Br(\bar{B}^0 \to \pi^{\pm}\rho^{\mp})$ agree 
with CLEO data\cite{CLEO} very well and
favor $\cos\gamma<0$ if their experimental center values are taken seriously. 
To meet its experimental center value and $\cos\gamma<0$, the decay 
$B^- \to \pi^- \omega$ will prefer larger form factor $A_{0}^{B\to\omega}(0)$.
For the other decay modes, the branching ratios are predicted well below their
$90\% C.L$ upper limits given in Ref\cite{CLEO}. We have also compared our results
with Cheng and Yang's renewed results of the branchig ratios. For many decay modes,
our results agree with theirs. 

Working in the QCD improved factorization approach, we are allowed to calculate
the strong phases to make predictions of CP asymetries for the decay modes more sound
than before. We find direct CP asymmetries in the observed decay modes 
 are around few percentages level. Direct and time-integrated CP asymmetries in
 those decay modes have been tabled in Fig.3 and Fig.4 respectively.


\section*{Acknowledgments}

We acknowledge the Grant-in-Aid for Scientific Research on Priority Areas
(Physics of CP violation with contract number 09246105 and 1014028) and the
Monbusho Found 10098178-00. We thanks JSPS(Japan Society for the Promotion of
Science) for support.

\newpage 
 
\begin{center} {\bf Appendix A}  \end{center}  

The decay amplitudes of some of the $B\to PV$ decays in terms of 
the effective coefficients $a_i$'s:
 
\begin{eqnarray} 
{\cal M}(\bar{B}^0 \to \pi^0 \rho^0 )=-\frac{G_{F}}{2\sqrt{2}} 
 f_{\rho}M_{B}^2 F_{0}^{B\to\pi}\lambda V_{cb}  
\biggl\{ 
           R_{u} e^{-i\gamma}  
      \left[ a_{2}(\bar{B}^0 \pi^{0},\rho^0 ) 
-a_{4}^{u}(\bar{B}^0 \pi^{0},\rho^0 ) 
      \right. 
 \biggr.&\nonumber \\ 
    \left. 
         +\frac{3}{2}\left( a_{7}(\bar{B}^0 \pi^{0},\rho^0 ) 
             +a_{9}(\bar{B}^0 \pi^{0},\rho^0 ) 
          \right)  
            +\frac{1}{2}a_{10}(\bar{B}^0 \pi^{0},\rho^0 ) 
            -a_{10a}(\bar{B}^0 \pi^{0},\rho^0 )  
    \right]  
&\nonumber \\ 
 \biggl. 
 -\left[ -a_{4}^{c}(\bar{B}^0 \pi^{0},\rho^0 ) 
         +\frac{3}{2}\left( a_{7}(\bar{B}^0 \pi^{0},\rho^0 ) 
           +a_{9}(\bar{B}^0 \pi^{0},\rho^0 )\right)  
  +\frac{1}{2}a_{10}(\bar{B}^0 \pi^{0},\rho^0 )-a_{10a}(\bar{B}^0 \pi^{0},\rho^0 ) 
 \right] 
   \biggr\} 
&\nonumber \\ 
   -\frac{G_{F}}{2\sqrt{2}}f_{\pi} M_{B}^2 A_{0}^{B\to \rho} \lambda V_{cb} 
 \biggl\{ 
     R_{u} e^{-i\gamma}  
    \left[ a_{2}(\bar{B}^0 \rho^{0},\pi^{0}) 
             -a_{4}^{u}(\bar{B}^0 \rho^{0},\pi^{0}) 
    \right.  
 \biggr. &\nonumber \\ 
 -\left( a_{6}^{u}(\bar{B}^0 \rho^{0},\pi^{0})-a_{8}(\bar{B}^0 \rho^{0},\pi^{0})/2 
                 +a_{8a}(\bar{B}^0 \rho^{0},\pi^{0})  
  \right) R_{\pi^0}- 
   \frac{3}{2} 
       \left( a_{7}(\bar{B}^0 \rho^{0},\pi^{0})-a_{9}(\bar{B}^0 \rho^{0},\pi^{0}) 
       \right) 
 &\nonumber \\ 
   \left. +\frac{1}{2}a_{10}(\bar{B}^0 \rho^{0},\pi^{0}) 
            -a_{10a}(\bar{B}^0 \rho^{0},\pi^{0}) 
   \right] 
&\nonumber \\ 
    -\left[ -a_{4}^{c}(\bar{B}^0 \rho^{0},\pi^{0})- 
      \left( a_{6}^{c}(\bar{B}^0 \rho^{0},\pi^{0})- 
       \frac{1}{2}a_{8}(\bar{B}^0 \rho^{0},\pi^{0})+ 
        a_{8a}(\bar{B}^0 \rho^{0},\pi^{0})  
       \right) R_{\pi^0} 
     \right. 
&\nonumber \\ 
\biggl.   
  \left. -\frac{3}{2}\left( a_{7}(\bar{B}^0 \rho^{0},\pi^{0}) 
       -a_{9}(\bar{B}^0 \rho^{0},\pi^{0})\right)  
      +\frac{1}{2}a_{10}\left( \bar{B}^0 \rho^{0},\pi^{0} 
                  \right)-a_{10a}(\bar{B}^0 \rho^{0},\pi^{0}) 
   \right]  
\biggr\};&    
\end{eqnarray} 
 
\begin{eqnarray} 
{\cal M}(B^- \to \pi^0 \rho^- )=\frac{G_{F}}{2} 
 f_{\rho}M_{B}^2 F_{0}^{B\to\pi}\lambda V_{cb} 
\biggl\{ 
        R_{u} e^{-i\gamma}  
   \left[ a_{1}(B^{-}\pi^{0},\rho^{-})  
          +a_{4}^{u}(B^{-}\pi^{0},\rho^{-})  
   \right. 
\biggr. &\nonumber \\ 
   \left.  
        +a_{10}(B^{-}\pi^{0},\rho^{-}) +a_{10a}(B^{-}\pi^{0},\rho^{-})  
   \right] 
 &\nonumber \\ 
 \biggl.   
   -\left[ a_{4}^{c}(B^{-}\pi^{0},\rho^{-}) +a_{10}(B^{-}\pi^{0},\rho^{-}) + 
           a_{10a}(B^{-}\pi^{0},\rho^{-}) 
     \right] 
 \biggr\}         
&\nonumber \\ 
   +\frac{G_{F}}{2} f_{\pi}M_{B}^2 A_{0}^{B\to \rho} \lambda V_{cb}  
\biggl\{ 
         R_{u} e^{-i\gamma}  
           \left[ a_{2}(\bar{B}^0 \rho^{-},\pi^{0}) 
                 -a_{4}^{u}(\bar{B}^0 \rho^{-},\pi^{0}) 
           \right. 
\biggr. 
&\nonumber \\ 
   -\left( a_{6}^{u}(\bar{B}^0 \rho^{-},\pi^{0})-\frac{1}{2}a_{8}(\bar{B}^0 \rho^{-},\pi^{0}) 
         +a_{8a}(\bar{B}^0 \rho^{-},\pi^{0})  
    \right) R_{\pi^0}- 
\frac{3}{2}\left( a_{7}(\bar{B}^0 \rho^{-},\pi^{0}) 
                -a_{9}(\bar{B}^0 \rho^{-},\pi^{0})  
           \right) 
&\nonumber \\ 
   \left. +\frac{1}{2}a_{10}(\bar{B}^0 \rho^{-},\pi^{0}) 
             -a_{10a}(\bar{B}^0 \rho^{-},\pi^{0}) 
    \right]  
&\nonumber \\ 
       -\left[ -a_{4}^{c}(\bar{B}^0 \rho^{-},\pi^{0}) 
             -\left( a_{6}^{c}(\bar{B}^0 \rho^{-},\pi^{0}) 
               -\frac{1}{2}a_{8}(\bar{B}^0 \rho^{-},\pi^{0})  
                      + a_{8a}(\bar{B}^0 \rho^{-},\pi^{0}) 
          \right) R_{\pi^0} 
        \right. 
 &\nonumber \\ 
\biggl. \left. 
   -\frac{3}{2}\left( a_{7}(\bar{B}^0 \rho^{-},\pi^{0}) 
    -a_{9}(\bar{B}^0 \rho^{-},\pi^{0})\right) 
    +\frac{1}{2}a_{10}(\bar{B}^0 \rho^{-},\pi^{0}) 
      -a_{10a}(\bar{B}^0 \rho^{-},\pi^{0}) 
         \right] 
 \biggr\};& 
\end{eqnarray} 
 
\begin{eqnarray} 
{\cal M}(\bar{B}^{0}\to \pi^{0}\omega)=-\frac{G_{F}}{2\sqrt{2}}  
f_{\omega} M_{B}^2 F_{0}^{B\to\pi}\lambda V_{cb}  
\biggl\{ 
      R_{u} e^{-i\gamma}  
    \left[ a_{2}(\bar{B}^{0}\pi^{0},\omega) 
           +a_{4}^{u}(\bar{B}^{0}\pi^{0},\omega) 
    \right. 
\biggr.  
&\nonumber \\ 
     +2\left( a_{3}(\bar{B}^{0}\pi^{0},\omega) 
     +a_{5}(\bar{B}^{0}\pi^{0},\omega)\right) 
      +\frac{1}{2} 
   \left( a_{7}(\bar{B}^{0}\pi^{0},\omega) 
          +a_{9}(\bar{B}^{0}\pi^{0},\omega) 
   \right. 
&\nonumber \\ 
   \left.  \left.   -a_{10}(\bar{B}^{0}\pi^{0},\omega) 
        +2 a_{10a}(\bar{B}^{0}\pi^{0},\omega) 
  \right) \right] 
&\nonumber \\ 
-\left[a_{4}^{c}(\bar{B}^{0}\pi^{0},\omega) 
        +2 \left( a_{3}(\bar{B}^{0}\pi^{0},\omega) 
           +a_{5}(\bar{B}^{0}\pi^{0},\omega) \right) 
  \right. 
&\nonumber \\ 
\biggl. \left.   
      +\frac{1}{2}\left( a_{7}(\bar{B}^{0}\pi^{0},\omega) 
        +a_{9}(\bar{B}^{0}\pi^{0},\omega)- 
        a_{10}(\bar{B}^{0}\pi^{0},\omega) 
             +2 a_{10a}\right) 
        \right] 
\biggr\} 
&\nonumber \\ 
+\frac{G_{F}}{2\sqrt{2}} f_{\pi}M_{B}^2 A_{0}^{B\to\omega } \lambda V_{cb}  
\biggl\{ 
           R_{u} e^{-i\gamma}  
         \left[ a_{2}(\bar{B}^{0}\omega,\pi^{0})-a_{4}^{u}(\bar{B}^{0}\omega,\pi^{0}) 
         \right. 
\biggr. 
&\nonumber \\ 
   -\left( a_{6}^{u}(\bar{B}^{0}\omega,\pi^{0}) 
          -\frac{1}{2}a_{8}(\bar{B}^{0}\omega,\pi^{0}) 
            +a_{8a}(\bar{B}^{0}\omega,\pi^{0}) 
    \right) R_{\pi^0} 
&\nonumber \\ 
  \left.   
  +\frac{3}{2}\left( a_{9}(\bar{B}^{0}\omega,\pi^{0})-a_{7}(\bar{B}^{0}\omega,\pi^{0}) 
                 \right)  
           +\frac{1}{2}a_{10}(\bar{B}^{0}\omega,\pi^{0})-a_{10a}(\bar{B}^{0}\omega,\pi^{0}) 
  \right] 
&\nonumber \\ 
        -\left[ -a_{4}^{c}(\bar{B}^{0}\omega,\pi^{0}) 
        -\left( a_{6}^{c}(\bar{B}^{0}\omega,\pi^{0})-\frac{1}{2}a_{8}(\bar{B}^{0}\omega,\pi^{0}) 
        +a_{8a}(\bar{B}^{0}\omega,\pi^{0}) 
       \right) R_{\pi^0} 
       \right. 
&\nonumber \\ 
   \biggl. \left. 
     +\frac{3}{2}\left( a_{9}(\bar{B}^{0}\omega,\pi^{0})- 
                        a_{7}(\bar{B}^{0}\omega,\pi^{0}) 
               \right) 
     +\frac{1}{2}a_{10}(\bar{B}^{0}\omega,\pi^{0})-a_{10a}(\bar{B}^{0}\omega,\pi^{0}) 
           \right] 
  \biggr\};& 
\end{eqnarray} 
 
\begin{eqnarray} 
{\cal M}(B^- \to K^{-}\rho^{0})= 
\frac{G_{F}}{2} f_{K} M_{B}^2 A_{0}^{B\to \rho} V_{cb} (1-0.5 \lambda^{2}) 
\biggl\{ 
       R_{c} e^{-i\gamma}  
    \left[ a_{1}(B^- \rho^{0},K^{-}) 
             +a_{4}^{u}(B^- \rho^{0},K^{-}) 
     \right. 
\biggr. 
&\nonumber \\ 
    \left. 
        +rki \left( a_{6}^{u}(B^- \rho^{0},K^{-})+a_{8}(B^- \rho^{0},K^{-}) 
                  +a_{8a}(B^- \rho^{0},K^{-}) 
           \right)+a_{10}(B^- \rho^{0},K^{-})+a_{10a}(B^- \rho^{0},K^{-}) 
    \right] 
&\nonumber \\ 
     +\left[a_{4}^{c}(B^- \rho^{0},K^{-}) 
       +rki \left( a_{6}^{c}(B^- \rho^{0},K^{-})+a_{8}(B^- \rho^{0},K^{-}) 
      +a_{8a}(B^- \rho^{0},K^{-}) 
           \right) 
       \right. 
&\nonumber \\ 
\biggl. \left.  +a_{10}(B^- \rho^{0},K^{-})+a_{10a}(B^- \rho^{0},K^{-}) 
        \right] 
\biggr\} 
&\nonumber \\ 
+\frac{G_{F}}{2} f_{\rho}M_{B}^2 F_{0}^{B\to K}V_{cb} (1-\frac{\lambda^2 }{2})  
\biggl\{ 
        R_{c} e^{-i\gamma}  
     \left[ a_{2}(B^{-}K^{-},\rho^{0}) 
                +\frac{3}{2}a_{7}(B^{-}K^{-},\rho^{0}) 
     \right. 
\biggr. 
&\nonumber \\ 
\biggl. 
        \left.  +\frac{3}{2}a_{9}(B^{-}K^{-},\rho^{0}) 
      \right] 
       +\frac{3}{2}\left[ a_{7}(B^{-}K^{-},\rho^{0})+a_{9}(B^{-}K^{-},\rho^{0}) \right] 
\biggr\};& 
\end{eqnarray} 
 
\begin{eqnarray} 
{\cal M}(B^- \to\pi^{-}K^{*})=\frac{G_{F}}{\sqrt{2}} f_{K^{*}} M_{B}^2  
           F_{0}^{B\to\pi^{-}} V_{cb} (1-\frac{\lambda^2 }{2})  
\biggl\{ 
        R_{c} e^{-i\gamma}  
      \left[ 
                a_{4}^{u}(B^{-}\pi^{-},K^{*}) 
                -\frac{1}{2}a_{10}(B^{-}\pi^{-},K^{*}) 
        \right. 
\biggr. 
&\nonumber \\ 
\biggl. \left.  
           +a_{10a}(B^{-}\pi^{-},K^{*}) 
       \right] 
     +\left[ a_{4}^{c}(B^{-}\pi^{-},K^{*})-\frac{1}{2}a_{10}(B^{-}\pi^{-},K^{*}) 
          +a_{10a}(B^{-}\pi^{-},K^{*}) 
      \right] 
\biggr\};& 
\end{eqnarray} 
 
\begin{eqnarray} 
{\cal M}(B^{-}\to K^{-}K^{*})=\frac{G_{F}}{\sqrt{2}} 
   f_{K^{*}} M_{B}^2 F_{0}^{B\to K}\lambda V_{cb}  
\biggl\{   
       R_{u} e^{-i\gamma} 
      \left[ 
         a_{4}^{u}(B^{-}K^{-},K^{*})-\frac{1}{2}a_{10}(B^{-}K^{-},K^{*}) 
       \right. 
\biggr. 
&\nonumber\\ 
\biggl. \left. 
        +a_{10a}(B^{-}K^{-},K^{*})   \right] 
        -\left[ a_{4}^{c}(B^{-}K^{-},K^{*})-\frac{1}{2}a_{10}(B^{-}K^{-},K^{*}) 
        +a_{10a}(B^{-}K^{-},K^{*}) 
          \right] 
\biggr\};& 
\end{eqnarray} 
 
\begin{eqnarray} 
{\cal M}(B^{-}\to\pi^{0}K^{*-})=\frac{G_{F}}{2} f_{K^{*}}  
              M_{B}^2 F_{0}^{B\to\pi}V_{cb} (1-\frac{\lambda^2 }{2})  
\biggl\{ 
         R_{c} e^{-i\gamma}  
    \left[ 
           a_{1}(B^{-}\pi^{0},K^{*-})+a_{4}^{u}(B^{-}\pi^{0},K^{*-}) 
    \right. 
\biggr. 
&\nonumber \\ 
  \left. 
         +a_{10}(B^{-}\pi^{0},K^{*-})+a_{10a}(B^{-}\pi^{0},K^{*-}) 
  \right] 
&\nonumber \\ 
 \biggl.  
  +\left[ 
       a_{4}^{c}(B^{-}\pi^{0},K^{*-})+a_{10}(B^{-}\pi^{0},K^{*-})+ 
                a_{10a}(B^{-}\pi^{0},K^{*-}) 
   \right] 
 \biggr\} 
&\nonumber \\ 
+\frac{G_{F}}{2} f_{\pi}M_{B}^2 A_{0}^{B\to K^{*}} V_{cb} (1-\frac{\lambda^2 }{2})  
 \biggl\{ 
        R_{c} e^{-i\gamma}  
     \left[ a_{2}(B^{-}K^{*-},\pi^{0})- \frac{3}{2}\left(a_{7}(B^{-}K^{*-},\pi^{0}) 
    \right.\right. 
\biggr. 
&\nonumber \\ 
\biggl. 
   \left. 
    -a_{9}(B^{-}K^{*-},\pi^{0}) 
   \right) 
   -\frac{3}{2}\left[a_{7}(B^{-}K^{*-},\pi^{0})-a_{9}(B^{-}K^{*-},\pi^{0})\right] 
\biggr\};& 
\end{eqnarray} 
 
\begin{eqnarray} 
{\cal M}(B^{-}\to K^{-}\omega)= 
\frac{G_{F}}{2} f_{\omega}  M_{B}^2 F_{0}^{B\to K}V_{cb} (1-\frac{\lambda^2 }{2})  
\biggl\{ 
      R_{c} e^{-i\gamma}  
    \left[ 
         a_{2}(B^{-}K^{-},\omega)+2 \left( a_{3}(B^{-}K^{-},\omega) 
                                    \right. 
    \right. 
\biggr. 
  &\nonumber \\ 
  \left.                        \left. 
                                      +a_{5}(B^{-}K^{-},\omega)  
                                   \right) 
    +\frac{1}{2}\left( a_{7}(B^{-}K^{-},\omega)+a_{9}(B^{-}K^{-},\omega)\right) 
  \right]   
&\nonumber \\ 
\biggl.       
     +\left[ 2 \left( a_{3}(B^{-}K^{-},\omega) 
        +a_{5}(B^{-}K^{-},\omega) \right) 
     +\frac{1}{2}\left( a_{7}(B^{-}K^{-},\omega)+a_{9}(B^{-}K^{-},\omega) 
                \right) 
        \right] 
\biggr\} 
&\nonumber \\ 
+\frac{G_{F}}{2} f_{K} M_{B}^2 A_{0}^{B\to\omega } V_{cb} (1-\frac{\lambda^2 }{2})  
\biggl\{  
        R_{c} e^{-i\gamma}  
       \left[ a_{1}(B^{-}\omega,K^{-}) 
              +a_{4}^{u} (B^{-}\omega,K^{-}) 
       \right. 
\biggr. 
&\nonumber \\ 
  \left. +R_{K^-} \left( a_{6}^{u}(B^{-}\omega,K^{-})+a_{8}(B^{-}\omega,K^{-}) 
        +a_{8a}(B^{-}\omega,K^{-}) 
      \right) 
       +a_{10}(B^{-}\omega,K^{-})+a_{10a}(B^{-}\omega,K^{-}) 
  \right] 
&\nonumber \\ 
+\left[ a_{4}^{c}(B^{-}\omega,K^{-}) 
        +R_{K^-} \left( a_{6}^{c}(B^{-}\omega,K^{-})+a_{8}(B^{-}\omega,K^{-}) 
                +a_{8a}(B^{-}\omega,K^{-}) 
             \right) 
 \right. 
&\nonumber \\ 
 \biggl. \left. 
    +a_{10}(B^{-}\omega,K^{-})+a_{10a}(B^{-}\omega,K^{-} 
   \right] 
    \biggr\};& 
\end{eqnarray} 
 
\begin{eqnarray} 
{\cal M}(\bar{B}^{0}\to \bar{K}^{0}\omega)=\frac{G_{F}}{2} f_{\omega}  M_{B}^2  
   F_{0}^{B\to K} V_{cb} (1-\frac{\lambda^2 }{2})  
 \biggl\{   
           R_{c} e^{-i\gamma}  
   \left[ a_{2}(\bar{B}^{0}\bar{K}^{0},\omega) 
         +2\left( a_{3}(\bar{B}^{0}\bar{K}^{0},\omega) 
           \right. 
   \right. 
\biggr.   
&\nonumber \\ 
   \left. \left.  
          +a_{5}(\bar{B}^{0} \bar{K}^{0},\omega) 
          \right) 
      +\frac{1}{2}\left( a_{7}(\bar{B}^{0}\bar{K}^{0},\omega) 
                  +a_{9}(\bar{B}^{0}\bar{K}^{0},\omega) 
                 \right) 
   \right] 
&\nonumber \\ 
\biggl.   
   +\left[ 2\left( a_{3}(\bar{B}^{0}\bar{K}^{0},\omega) 
         +a_{5}(\bar{B}^{0}\bar{K}^{0},\omega) 
            \right) 
        +\frac{1}{2}\left( a_{7}(\bar{B}^{0}\bar{K}^{0},\omega) 
                         +a_{9}(\bar{B}^{0}\bar{K}^{0},\omega) 
                    \right) 
     \right] 
\biggr\}         
&\nonumber \\ 
+\frac{G_{F}}{2} f_{K} M_{B}^2 A_{0}^{B\to\omega } V_{cb} (1-\frac{\lambda^2 }{2})  
\biggl\{ 
      R_{c} e^{-i\gamma} 
    \left[ 
         a_{4}^{u}(\bar{B}^{0}\omega,\bar{K}^{0}) 
       +R_{K^0} \left(a_{6}^{u}(\bar{B}^{0}\omega,\bar{K}^{0}) 
            \right. 
    \right.  
\biggr. 
 &\nonumber \\ 
           \left.   -\frac{1}{2}a_{8}(\bar{B}^{0}\omega,\bar{K}^{0}) 
                        +a_{8a}(\bar{B}^{0}\omega,\bar{K}^{0}) 
            \right) 
 &\nonumber \\ 
    \left.    
       -\frac{1}{2}a_{10}(\bar{B}^{0}\omega,\bar{K}^{0}) 
          +a_{10a}(\bar{B}^{0}\omega,\bar{K}^{0}) 
    \right. 
    +\left[ a_{4}^{c}(\bar{B}^{0}\omega,\bar{K}^{0}) 
     \right.  
&\nonumber \\ 
\biggl. \left. 
+R_{K^-} \left( a_{6}^{c}(\bar{B}^{0}\omega,\bar{K}^{0}) 
          -\frac{1}{2}a_{8}(\bar{B}^{0}\omega,\bar{K}^{0}) 
       +a_{8a}(\bar{B}^{0}\omega,\bar{K}^{0}) 
      \right) 
-\frac{1}{2}a_{10}(\bar{B}^{0}\omega,\bar{K}^{0})+a_{10a}(\bar{B}^{0}\omega,\bar{K}^{0}) 
       \right] 
 \biggr\};& 
\end{eqnarray} 
 
\begin{eqnarray} 
{\cal M}(\bar{B}^{0}\to K^{-}\rho^{+})=\frac{G_{F}}{\sqrt{2}}  
fk M_{B}^2 A_{0}^{B\to \rho} V_{cb} (1-\frac{\lambda^2 }{2})  
\biggl\{ 
          R_{c} e^{-i\gamma}  
      \left[ a_{1}(\bar{B}^0 \rho^{+},K^{-}) 
               +a_{4}^{u}(\bar{B}^0 \rho^{+},K^{-}) 
       \right. 
\biggr. 
&\nonumber \\ 
+R_{K^-} \left( a_{6}^{u}(\bar{B}^0 \rho^{+},K^{-})+a_{8}(\bar{B}^0 \rho^{+},K^{-}) 
        +a_{8a}(\bar{B}^0 \rho^{+},K^{-}) 
      \right) 
&\nonumber \\ 
   \left.  
    +a_{10}(\bar{B}^0 \rho^{+},K^{-})+a_{10a}(\bar{B}^0 \rho^{+},K^{-}) 
    \right] 
   +\left[a_{4}^{c}(\bar{B}^0 \rho^{+},K^{-}) 
     \right. 
&\nonumber \\ 
\biggl. \left. 
   +R_{K^-}\left(a_{6}^{c}(\bar{B}^0 \rho^{+},K^{-})+a_{8}(\bar{B}^0 \rho^{+},K^{-}) 
                +a_{8a}(\bar{B}^0 \rho^{+},K^{-}) 
    \right) 
   +a_{10}(\bar{B}^0 \rho^{+},K^{-})+a_{10a}(\bar{B}^0 \rho^{+},K^{-}) 
   \right] 
\biggr\}  ;& 
\end{eqnarray} 
 
\begin{eqnarray} 
{\cal M}(\bar{B}^{0}\to \pi^{0}\bar{K}^{*0}   )= 
\frac{G_{F}}{2} f_{\pi}M_{B}^2 A_{0}^{B\to K^{*}} V_{cb} (1-\frac{\lambda^2 }{2})  
 \biggl\{  
        R_{c} e^{-i\gamma} 
     \left[  
          a_{2}(\bar{B}^{0}\bar{K}^{*0}   ,\pi^{0})- 
        \frac{3}{2}\left( a_{7}(\bar{B}^{0}\bar{K}^{*0}   ,\pi^{0}) 
                   \right. 
     \right. 
 \biggr.  
&\nonumber \\ 
    \left. 
        -a_{9}(\bar{B}^{0}\bar{K}^{*0}   ,\pi^{0}) 
    \right) 
-\frac{3}{2}\left[ a_{7}(\bar{B}^{0}\bar{K}^{*0}   ,\pi^{0})-a_{9}(\bar{B}^{0}\bar{K}^{*0}   ,\pi^{0}) 
            \right]  
&\nonumber \\ 
-\frac{G_{F}}{2} f_{K^{*}} M_{B}^2 F_{0}^{B\to\pi}V_{cb} (1-\frac{\lambda^2 }{2})  
  \biggl\{ 
         R_{c} e^{-i\gamma} 
         \left[ 
                        a_{4}^{u}(\bar{B}^{0}\pi^{0},\bar{K}^{*0}   ) 
         -\frac{1}{2}a_{10}(\bar{B}^{0}\pi^{0},\bar{K}^{*0}   ) 
         \right. 
 \biggr. 
&\nonumber \\ 
  \biggl. \left.  
            +a_{10a}(\bar{B}^{0}\pi^{0},\bar{K}^{*0}   )    
          \right] 
  +\left[  
     a_{4}^{c}(\bar{B}^{0}\pi^{0},\bar{K}^{*0}   )-\frac{1}{2}a_{10}(\bar{B}^{0}\pi^{0},\bar{K}^{*0}   ) 
     +a_{10a}(\bar{B}^{0}\pi^{0},\bar{K}^{*0}   ) 
   \right] 
 \biggr\} ;& 
\end{eqnarray}

\newpage 
 
\begin{center} {\bf Figure captions} \end{center} 
 
\small{Fig.1 Order $\alpha_s$  non-factorizable contributions 
in $B\to M_1 M_2$ decays} 
 
\small{Fig.2 $BR( B\to PV)$ as a function of $\gamma$ are shown as  
 curves in units of $10^{-6}$. The BR measured by CLEO  
Collaboration are shown by horizontal solid lines. The thicker solid lines are  
its center values, thin lines are its error bars or the upper limit.} 
 
\small{Fig.3 Direct $CP$ asymmetry of $B\to PV$ as a function of $\gamma$ } 
 
\small{Fig.4 Time-integrated $CP$ asymmetry as a function of $\gamma$ with 
             the mixing parameter $\Delta m/\Gamma=0.723$ in the SM }

\newpage 
 
\begin{figure}[htbp] 
 \scalebox{0.7}{ 
 {\color{Red} 
 \fbox{\color{Black} 
   \begin{picture}(140,120)(-30,0) 
    \ArrowLine(0,40)(30,40) 
    \ArrowLine(30,40)(60,40) 
    \ArrowLine(60,40)(90,40)   
    \ArrowLine(90,20)(0,20) 
    \Gluon(30,40)(46,87){4}{4} \Vertex(30,40){1.5} \Vertex(46,87){1.5}  
    \Line(58,42)(62,38) 
    \Line(58,38)(62,42) 
    \ArrowLine(40,105)(60,45) 
    \ArrowLine(60,45)(80,105) 
    \put(-20,28){$\bar{B}$} 
    \put(90,28){$M_{1}$} 
    \put(58,110){$M_{2}$} 
    \put(55,28){\small{${\cal O}_i $}} 
    \put(0,45){\small{$b$}} 
    \put(45,0){(a)} 
 \end{picture} 
 }}} 
 \scalebox{0.7}{ 
 {\color{Red} 
 \fbox{\color{Black} 
   \begin{picture}(140,120)(-30,0) 
    \ArrowLine(0,40)(30,40) 
    \ArrowLine(30,40)(60,40) 
    \ArrowLine(60,40)(90,40)   
    \ArrowLine(90,20)(0,20) 
    \Gluon(30,40)(74,87){4}{6} \Vertex(30,40){1.5} \Vertex(74,87){1.5}  
    \Line(58,42)(62,38) 
    \Line(58,38)(62,42) 
    \ArrowLine(40,105)(60,45) 
    \ArrowLine(60,45)(80,105) 
    \put(-20,28){$\bar{B}$} 
    \put(90,28){$M_{1}$} 
    \put(58,110){$M_{2}$} 
    \put(55,28){\small{${\cal O}_i $}} 
    \put(0,45){\small{$b$}} 
    \put(45,0){(b)} 
 \end{picture} 
 }}} 
 \scalebox{0.7}{ 
 {\color{Red} 
 \fbox{\color{Black} 
   \begin{picture}(140,120)(-30,0) 
    \ArrowLine(0,40)(20,40) 
    \ArrowLine(20,40)(60,40) 
    \ArrowLine(60,40)(90,40)   
    \ArrowLine(90,20)(0,20) 
    \Gluon(54,87)(65,40){3}{6} \Vertex(65,40){1.5} \Vertex(54,87){1.5}  
    \Line(38,42)(42,38) 
    \Line(38,38)(42,42) 
    \ArrowLine(20,105)(40,45) 
    \ArrowLine(40,45)(60,105) 
    \put(-20,28){$\bar{B}$} 
    \put(90,28){$M_{1}$} 
    \put(58,110){$M_{2}$} 
    \put(35,28){\small{${\cal O}_i $}} 
    \put(0,45){\small{$b$}} 
    \put(45,0){(c)} 
 \end{picture} 
 }}} 
 \scalebox{0.7}{ 
 {\color{Red} 
 \fbox{\color{Black} 
   \begin{picture}(140,120)(-30,0) 
    \ArrowLine(0,40)(20,40) 
    \ArrowLine(20,40)(60,40) 
    \ArrowLine(60,40)(90,40)   
    \ArrowLine(90,20)(0,20) 
    \Gluon(26,87)(65,40){3}{6} \Vertex(65,40){1.5} \Vertex(54,87){1.5}  
    \Line(38,42)(42,38) 
    \Line(38,38)(42,42) 
    \ArrowLine(20,105)(40,45) 
    \ArrowLine(40,45)(60,105) 
    \put(-20,28){$\bar{B}$} 
    \put(90,28){$M_{1}$} 
    \put(58,110){$M_{2}$} 
    \put(35,28){\small{${\cal O}_i $}} 
    \put(0,45){\small{$b$}} 
    \put(45,0){(d)} 
 \end{picture} 
 }}} 
 
\vspace{1cm}  
\scalebox{0.7}{ 
 {\color{Red} 
 \fbox{\color{Black} 
   \begin{picture}(140,120)(-30,0) 
    \ArrowLine(90,20)(-5,20) 
    \ArrowLine(-5,50)(20,50) 
    \ArrowLine(20,50)(30,100) 
    \ArrowLine(70,50)(90,50)   
    \ArrowLine(80,100)(70,50) 
   \put(35,50){\circle{27}} 
    \Gluon(50,50)(70,50){3}{3} \Vertex(50,50){1.5} \Vertex(70,50){1.5}  
    \Line(18,52)(22,48) 
    \Line(18,48)(22,52) 
    \put(-20,28){$\bar{B}$} 
    \put(90,28){$M_{1}$} 
    \put(58,110){$M_{2}$} 
    \put(23, 45){\small{${\cal O}_i $}} 
    \put(0,54){\small{$b$}} 
    \put(45,0){(e)} 
 \end{picture} 
 }}} 
\scalebox{0.7}{ 
 {\color{Red} 
 \fbox{\color{Black} 
   \begin{picture}(140,120)(-30,0) 
    \ArrowLine(90,20)(-5,20) 
    \ArrowLine(-5,50)(20,50) 
    \ArrowLine(20,50)(30,100) 
    \ArrowLine(70,50)(90,50)   
    \ArrowLine(80,100)(70,50) 
    \Gluon(20,50)(70,50){3}{6} \Vertex(20,50){1.5} \Vertex(70,50){1.5}  
    \Line(18,52)(22,48) 
    \Line(18,48)(22,52) 
    \put(-20,28){$\bar{B}$} 
    \put(90,28){$M_{1}$} 
    \put(58,110){$M_{2}$} 
    \put(19, 40){\small{${\cal O}_g $}} 
    \put(0,54){\small{$b$}} 
    \put(45,0){(f)} 
 \end{picture} 
 }}} 
\scalebox{0.7}{ 
 {\color{Red} 
 \fbox{\color{Black} 
   \begin{picture}(140,120)(-30,0) 
    \ArrowLine(0,40)(60,40) 
    \ArrowLine(60,40)(90,40)   
    \ArrowLine(90,20)(0,20) 
    \Gluon(30,20)(46,87){4}{10} \Vertex(30,20){1.5} \Vertex(46,87){1.5}  
    \Line(58,42)(62,38) 
    \Line(58,38)(62,42) 
    \ArrowLine(40,105)(60,45) 
    \ArrowLine(60,45)(80,105) 
    \put(-20,28){$\bar{B}$} 
    \put(90,28){$M_{1}$} 
    \put(58,110){$M_{2}$} 
    \put(55,28){\small{${\cal O}_i $}} 
    \put(0,45){\small{$b$}} 
    \put(45,0){(g)} 
 \end{picture} 
 }}} 
\scalebox{0.7}{ 
 {\color{Red} 
 \fbox{\color{Black} 
   \begin{picture}(140,120)(-30,0) 
    \ArrowLine(0,40)(20,40) 
    \ArrowLine(20,40)(90,40)   
    \ArrowLine(90,20)(0,20) 
    \Gluon(54,87)(65,20){3}{6} \Vertex(65,20){1.5} \Vertex(54,87){1.5}  
    \Line(38,42)(42,38) 
    \Line(38,38)(42,42) 
    \ArrowLine(20,105)(40,45) 
    \ArrowLine(40,45)(60,105) 
    \put(-20,28){$\bar{B}$} 
    \put(90,28){$M_{1}$} 
    \put(58,110){$M_{2}$} 
    \put(35,28){\small{${\cal O}_i $}} 
    \put(0,45){\small{$b$}} 
    \put(45,0){(h)} 
 \end{picture} 
 }}} 
\caption{ } 
\end{figure} 
 
\newpage 
 
\begin{figure} 
\scalebox{1.0}{\epsfig{file=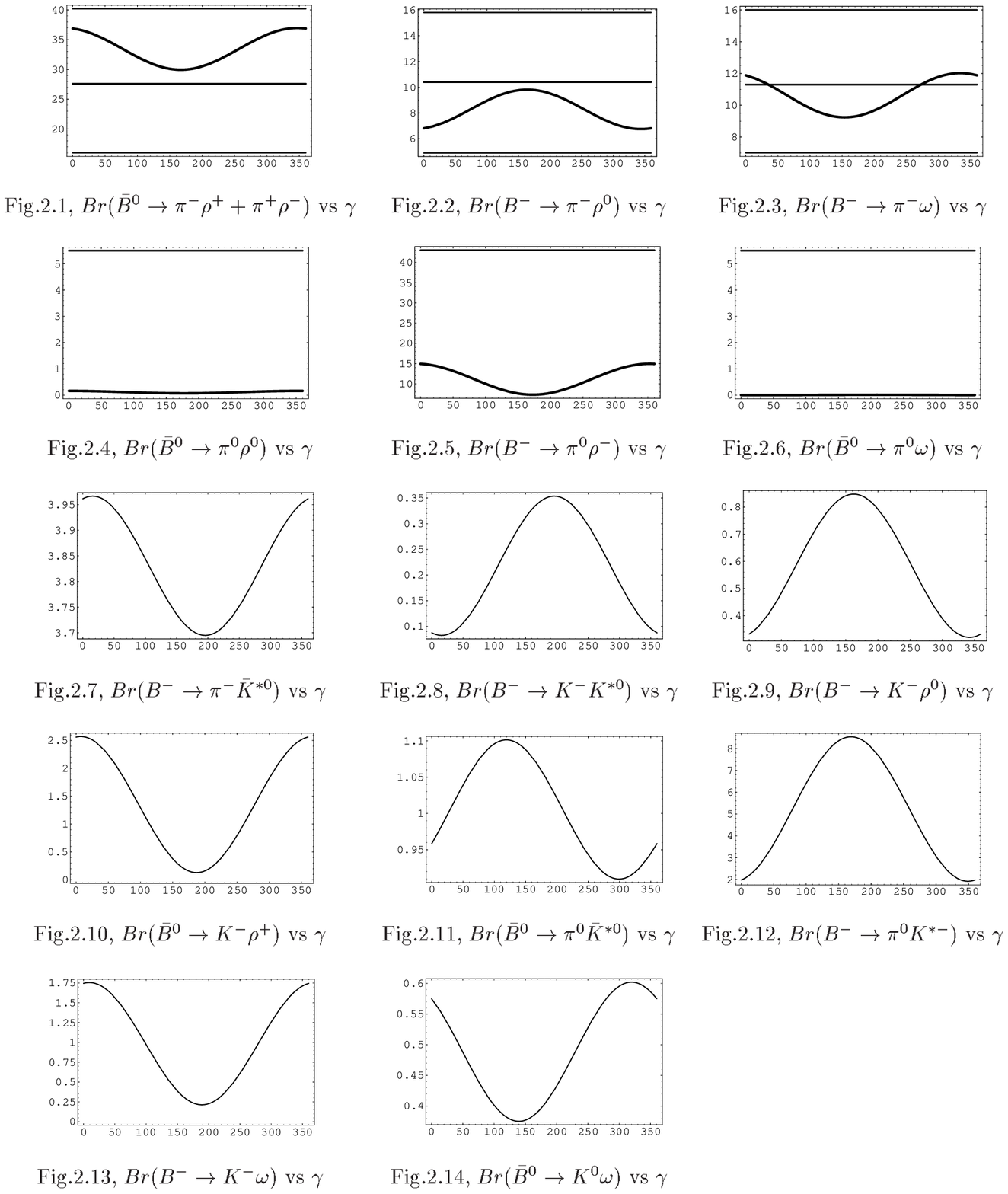}} 
\caption{}     
\end{figure} 
 
\newpage

\begin{figure} 
\scalebox{1.0}{\epsfig{file=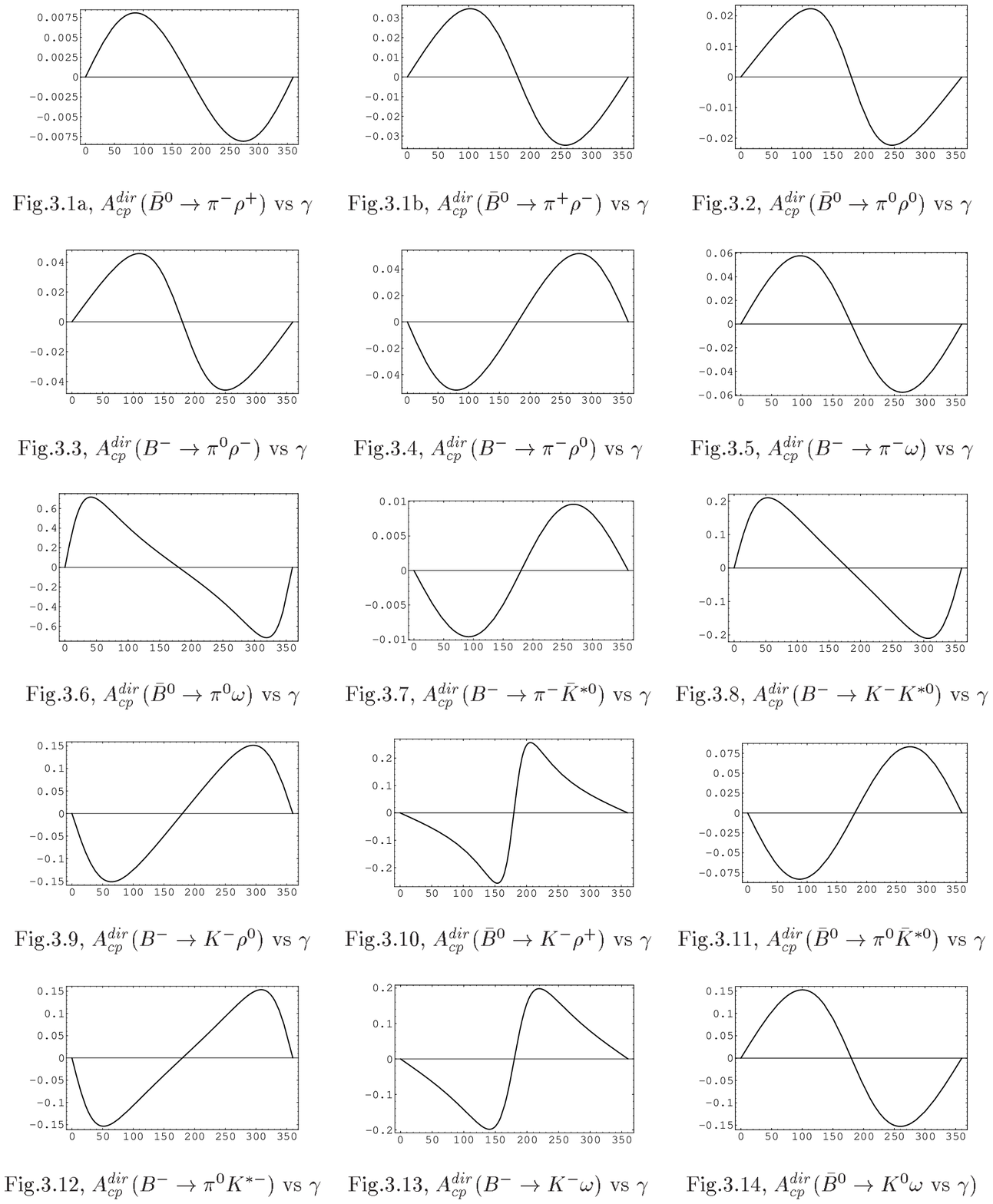}} 
\caption{}     
\end{figure} 
 
\newpage

\begin{figure} 
\begin{center} \scalebox{1.0}{\epsfig{file=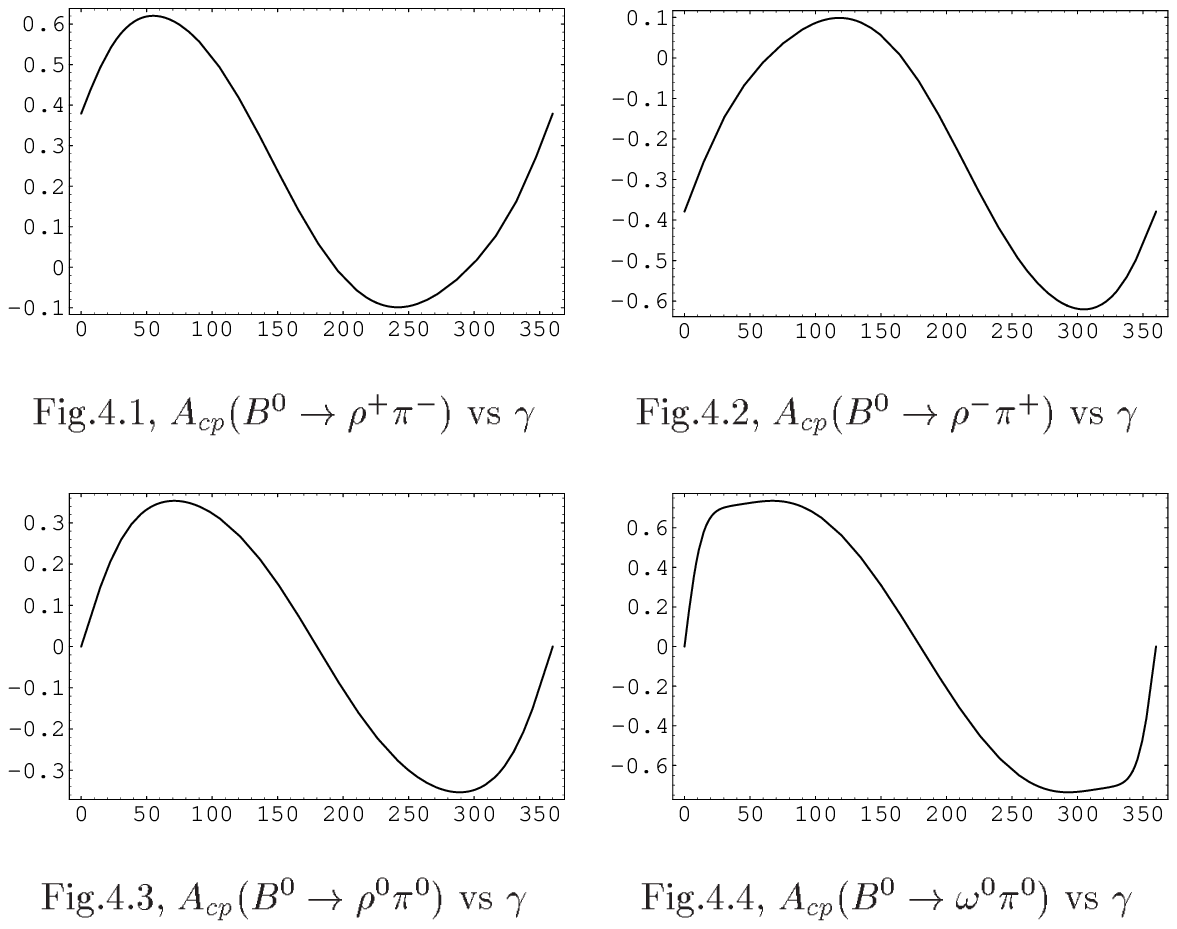}} \end{center} 
\caption{}     
\end{figure}

\end{document}